\documentclass[preprint,showpacs,prb,preprintnumbers,amsmath,amssymb]{revtex4}

\usepackage{graphicx}
\usepackage{dcolumn}
\usepackage{bm}

\begin{document}
\title{The role of grain boundaries in determining the transport properties of magnetite}

\author{D.C. Mertens$^1$, W. Montfrooij$^1$, R.J. McQueeney$^2$, M. Yethiraj$^3$, and J.M. Honig$^4$}

\affiliation{$^1$ Department of Physics and Astronomy, University of Missouri, Columbia, MO 65211\\
$^2$ Department of Physics and Astronomy, Iowa State University/Ames Laboratory, Ames, IA\\
$^3$ Center for Neutron Scattering, Oak Ridge National Laboratory, Oak Ridge, TN\\
$^4$ Chemistry Department, Purdue University, West Lafayette, IN}

\begin{abstract}
{We present magnetoresistance and Hall-effect measurements on single crystal magnetite (Fe$_3$O$_4$) close to the Verwey transition $T_V$= 123.8 K. We show that the formation of grain boundaries accompanying the reduction in crystal symmetry plays a significant role in the electron scattering mechanism, and that grain boundaries can account for the apparent change in sign of the charge carriers below $T_V$.} 
\end{abstract}
\maketitle

Despite extensive studies spanning seven decades\cite{garcia}, magnetite continues to attract considerable interest. Ever since the proposal by Verwey\cite{verwey} to attribute the two orders of magnitude increase in resistance, observed on cooling below $T\sim$124 K, in terms of a metal-insulator transition driven by charge ordering, a multitude of studies have been carried out\cite{garcia} in order to resolve the electronic structure of the charge ordered (CO) phase. However, there still does not exist a consensus on the low temperature phase of magnetite: for instance, the sign of the charge carriers below $T_V$ has not been established\cite{garcia}, the CO-pattern has not been determined from experiment, the concept of the metal-insulator transition has come under fire\cite{park} and more recently, the very idea of charge ordering has been challenged\cite{garcia}.\\

The properties of the CO-phase have been shown to depend strongly on the purity and stoichiometry of the sample\cite{aragon,aragon2,shepherd}. In view of the range of results presented for natural and synthetic magnetite crystals, and in view of the questions regarding the existence of a CO-state, we have opted to repeat standard measurements (specific heat $c_p$, resistance $R$, magnetoresistance $R_M$, magnetization $M$, Hall-effect $R_H$ and ac susceptibility) on synthetic, high-purity single crystals. We have combined these macroscopic measurements with elastic and inelastic polarized neutron scattering experiments that probe the microscopic nuclear and magnetic response. Some of the neutron scattering results are presented elsewhere in these proceedings\cite{mcqueeney}, in here we focus on the transport measurements.\\ 

The measurements were taken on two synthetic single crystal platelets\cite{david} of 31.5 and 24.3 mg. The orientation of the 24.3 mg crystal and its mosaicity (6') used in the acquisition of direction dependent magnetization data were determined by means of X-ray scattering. (Magneto)resistance measurements were done on the non-oriented sample. The transport and magnetization measurements were carried out using a Quantum Design PPMS-9 system in a temperature $T$ range of 2 $<$ $T$ $<$ 340 K, and for magnetic fields $H$ in the range -9 $<$ $H$ $<$ 9 T. The resistance measurements  were taken using a four probe method with the leads soldered or silver-pasted to the sample, the latter contacts being more resilient to thermal and magnetic cycling. Using pasted contacts lead to an increased resistance above $T_V$, reducing the jump in resistance at $T_V$ to $\sim$ 0.4 k$\Omega$ (Fig. \ref{tv}a). Below $T_V$ the two methods yield similar results, albeit displaying slightly different activation energies, possibly reflecting the difference in applied current direction. The change in Hall voltage on going through $T_V$ was similar for both sets of leads.\\

All our data demonstrate that our crystals are of high quality. We show typical measurements near the CO-phase transition in Fig. \ref{tv}. We observe an order of magnitude change in $R$ at $T_V$= 123.8 K in zero field (Fig. \ref{tv}a). The specific heat data clearly shows that the transition is first order (Fig. \ref{tv}b). We also show the change in magnetization for $H$= 0.01 T (100 Oe) along the (001) direction (Fig. \ref{tv}c). A large change in magnetization is a common observation in magnetite at low fields\cite{ziese}. This change in $M$ decreases with increasing $H$: it is barely visible at $H$= 1 T, and is absent in higher fields\cite{david}. Saturation magnetization is achieved around 1 T. The sharpness of the transition is also evident in the imaginary part of the ac-susceptibility at a frequency of 500 Hz (Fig. \ref{tv}d). Thermoremnance measurements for 4 $<T<$ 340 K did not reveal an impurity phase. All methods yield very similar transition temperatures (Fig. \ref{tv}), attesting\cite{aragon,aragon2,shepherd} to a homogenous oxygen distribution, with very small oxygen deficiency.\\

Application of a magnetic field lowers the Verwey transition temperature. Ziese and Blythe showed\cite{ziese}, using magnetoresistance measurements, that a magnetic field in the direction of the applied current lowered the transition temperature by $\sim$ 0.5 K for $H$= 1 T; no effect was observed\cite{ziese} for fields perpendicular to the current. We find a small shift ($\sim$ 0.3 K) in applied fields of 9 T when $R_H$ is used as a probe. On cooling down, a jump in $R_H$ is observed at $T_V$. This jump can also be invoked by applying a magnetic field at constant $T$ (Fig. \ref{tv_vs_h}a), leading to the observed field dependence of $T_V$ shown in the inset of Fig. \ref{tv_vs_h}a (corrected for the known field dependence of the temperature control). It is unclear from these data whether this represents a change in response to a perpendicular field or merely a change in response associated with a longitudinal field in conjunction with a small sample tilt ($\sim$ 3 degrees). Measurements are in progress\cite{david} to settle this issue.\\

Notwithstanding the crystal passing all purity checks, we also find that we can reproduce some of the more puzzling features that have been observed in magnetoresistance and Hall-effect measurements (such as a change in sign\cite{kuipers} of the charge carriers in the CO-phase). We argue that these features are related to the formation of grain boundaries\cite{coey} inherent to the twinning that accompanies the change\cite{iizumi} from a cubic spinel structure to a monoclinic phase. Magnetoresistance measurements\cite{ziese,kostopoulos,gridin} probe the role that grain boundaries play in the electron scattering process. The importance of grain boundaries in the electron scattering mechanism has already been established in thin film and powder studies\cite{coey}. The morphology of the grain boundaries is determined by the thermal and magnetic history of the sample. It is known that a moderate magnetic field applied along one of the cubic axes while cooling through $T_V$ is sufficient for establishing that cubic axis as the c$^{*}$-direction in the ordered phase. Of course, this does not prevent twinning. Magnetic fields in excess of 1 T ensure that there are no magnetic domain boundaries, but it is not entirely clear how magnetic fields affect the grain boundaries through a crystal axis switching\cite{vittoratos} mechanism. However, the presence of grain boundaries strongly influences the electron scattering mechanism and therefore, magnetoresistance measurements give insight into the interplay between grain boundary formation and magnetic field. Finally, once the c$^{*}$-axis has been established, the field can be released, resulting in a macroscopically demagnetized multi-domain sample, but with unchanged c$^{*}$-axis.\\

We measured the magnetoresistance $R_M$ for fields applied perpendicular to the current direction. Just below $T_V$ this results in a very rapid decrease in resistance for small $H$, followed by a much smaller decrease for larger $H$. Given the field dependence of $T_V$, this initial rapid decrease\cite{gridin} should be interpreted as the sample being driven from the ordered into the disordered state. This interpretation is consistent with the changes observed in $R_H$. Further below $T_V$, this very rapid decrease is no longer observed (see inset Fig. \ref{tv_vs_h}b). Instead, repeated cycling in $H$ at constant $T$ shows that $R_M$ values differ somewhat when revisiting the same point in the $H-T$ phase diagram, reflecting the changed morphology of the grain boundaries due to axis switching. Above $T_V$, where crystal axis switching does not take place\cite{vittoratos}, one simply observes the influence of magnetic domain walls on $R_M$; cycling in $H$ reproduces the same value when the same point in $H-T$ space is revisited.\\

Kuipers and Brabers\cite{kuipers} have analyzed thermoelectric-power measurements on pure and doped magnetite crystals, and concluded that minor changes in stoichiometry have a profound influence on whether $p$ or $n$-type conduction is observed below $T_V$. Here we show that the formation of grain boundaries accompanying the monoclinic distortion below $T_V$ can mimic the effects of changes in stoichiometry. In particular, at a grain boundary the ratio Fe$^{2+}$/Fe$^{3+}$ will be affected, effectively introducing a vacancy concentration $V_{\gamma}$. Therefore, one can expect to observe similar changes\cite{kuipers} in sign of charge carriers below $T_V$ as observed in thermoelectric-power measurements where the results were modeled based on the modified chemical formula Fe$^{3+}_{2+2\gamma}$ Fe$^{2+}_{1-3\gamma}$V$_{\gamma}$O$_4$.\\

To demonstrate this effect, we measured $R_H$ below $T_V$ for various preparation histories. As mentioned, the exact (H-T) preparation affects the grain boundary size and distribution. In agreements with thermoelectric-power measurements\cite{kuipers}, we observe a steep change in $R_H$ on cooling through the transition (Fig. \ref{tv_vs_h}a). It is tempting to ascribe this to a change in number of charge carriers, however, the $T$ and $H$ dependence of $R_H$ below $T_V$ show the issue to be more complicated. In Fig. \ref{hall} we show $R_H(T)$ for three preparation histories. Clearly, the datasets show a distinctly different behavior (Fig. \ref{hall}a), even though $R_M$ values remained virtually unchanged (Fig. \ref{hall}b). We find that we can induce a change of sign in $R_H$ similar to the change in sign in Seebeck coefficient observed in the thermoelectric-power measurements\cite{kuipers}. Given the dependence of the observed effect on preparation history, we do not believe that this change in sign should be attributed to an intrisic property, rather it should be attributed to grain boundary effects.\\

In conclusion, we have shown that the formation of grain boundaries, inherent to a low temperature structure with small monoclinic angle\cite{iizumi}, strongly influences the results of magnetoresistance measurements. It also renders interpretation of Hall-effect measurements nigh impossible as the sign of the Hall voltage depends on the morphology of the grain boundaries. In fact, in addition to the increase in activation energy\cite{park}, charge localization at the grain boundaries might play some part in the jump in resistance at $T_V$. Whether such a speculation could provide a (partial) explanation for the observations by Garcia {\it et al.} that charge order does not take place\cite{garcia} requires further study.\\ 

We thank Dr. P. Miceli for his assistance with the X-ray characterization and for the usage of his equipment.

\begin{figure} [b]
\caption{The Verwey transition in Fe$_3$O$_4$ probed using four different methods. (a) The resistance $R$ drops sharply at $T=T_V$ = 123.8 K (dotted line, soldered sample; solid line, silver pasted sample). We also show the expected lineshape for thermally activated conduction $R \sim exp[U/T]$ ($U$ = 100 meV, dashed line). (b) The specific heat measurements demonstrate that the transition is first order. The scatter in points at $T_V$ results from the breakdown of the heat pulse method near $T_V$. (c) The magnetization $M$ shows a jump at $T_V$ when a small field (0.01 T) is applied along the (001)-direction. (d) The imaginary part of the ac-susceptibility shows the enhanced dissipation close to $T_V$.}\label{tv}

\end{figure}

\begin{figure} [b]
\caption{The effect of applied field on resistance $R_M$ and Hall-resistance $R_H$ near $T_V$ at $T$= 123.6 K. The numbers and arrows in (a) and (b) correspond to the order of field cycling. (a) The transition into the ordered phase is marked by a sharp change in $R_H$. The system can be driven into the disordered phase by application of a magnetic field, leading to the $H$-dependent $T_V$ shown in the inset. (b) A large change in $R_M$ just below $T_V$ is observed when $H$ is applied perpendicular to the current direction. The magnitude of the change in $R_M$ corresponds to the measured (Fig. \ref{tv}a) zero-field change in $R$ at $T_V$, and therefore it should be interpreted as the system being driven into the disordered state. The inset shows the behavior of $R_M$ with field cycling further above $T_V$ (top panel, $T$= 130 K) and below (bottom panel, $T$= 120 K).} \label{tv_vs_h}
\end{figure}

\begin{figure} [b]
\caption{When $T$ is lowered through $T_V$, $R_H$ depends strongly on the preparation history. (a) We show three Hall measurements for applied fields of -7 T, -9 T and +9 T; all measurements were taken by cooling down from 130 K. As can be seen, the formation of grain boundaries associated with the monoclinic phase can lead to vastly different $R_H$ values, and can even change the sign of $R_H$. (b) However, this effect is not observed in $R_M$ data taken simultaneously with the curves in part (a).}\label{hall}
\end{figure}

\begin{thebibliography}{10}
\bibitem{garcia} Joaquin Garcia and Gloria Subias, J. Phys.: Condens. Matter 16, R145 (2004), and references therein.
\bibitem{verwey} E.J.W. Verwey and P.W. Haayman, Physica 8, 979 (1941).
\bibitem{park} J.-H. Park {\it et al.}, Phys. Rev. B55, 12813 (1997).
\bibitem{aragon} R. Aragon {\it et al.}, Phys. Rev. B31, 430 (1985).
\bibitem{aragon2} R. Aragon, P.M. Gehring and S.M. Shapiro, Phys. Rev. Lett. 70, 1635 (1993).
\bibitem{shepherd} J.P. Shepherd {\it et al.}, Phys. Rev. B43, 8461 (1991).
\bibitem{mcqueeney} R.J. McQueeney {\it et al.}, in these proceedings.
\bibitem{david} D.C. Mertens {\it et al.}, to be published.
\bibitem{ziese} M. Ziese and H.J. Blythe, J. Phys.: Condens. Matter 12, 13 (2000).
\bibitem{kuipers} A.J.M Kuipers and V.A.M. Brabers, Phys. Rev. B14, 1401 (1976); Phys. Rev. B20, 594 (1979).
\bibitem{coey} J.M.D. Coey {\it et al.}, Appl. Phys. Lett. 72, 734 (1998).
\bibitem{iizumi} M. Iizumi {\it et al.}, Acta Cryst. B38, 2121 (1982). 
\bibitem{kostopoulos} D. Kostopoulos and K. Alexopoulos, J. Appl. Phys. 47, 1714 (1976).
\bibitem{gridin} V.V. Gridin, G.R. Hearne and J.M. Honig, Phys. Rev. B53, 15518 (1996). 
\bibitem{vittoratos} E. Vittoratos, I. Baranov and P.P.M. Meincke, J. Appl. Phys. 42, 1633 (1971). 

\end{thebibliography}
\end{document}